\documentclass[twocolumn,aps,showpacs,prb,superscriptaddress,tightenlines,amsmath,amssymb]{revtex4}
\usepackage{graphicx}
\usepackage{amssymb}
\usepackage{dcolumn}
\usepackage{amsmath}
\usepackage{bm}

\usepackage{colordvi}

\begin{document}

\title{Electron spin relaxation in $n$-type InAs quantum wires}

\author{C. L\"u}
\affiliation{Hefei National Laboratory for Physical Sciences at
Microscale, University of Science and Technology of China, Hefei,
Anhui, 230026, China}
\affiliation{Department of Physics, University
of Science and Technology of China, Hefei, Anhui, 230026, China}
\author{H. C. Schneider}
\affiliation{Physics Department and Research Center OPTIMAS,
  University of Kaiserslautern, 67653 Kaiserslautern, Germany}
\author{M.\ W.\ Wu}
\thanks{Author to  whom correspondence should be addressed}
\email{mwwu@ustc.edu.cn.}
\affiliation{Hefei National Laboratory for Physical Sciences at
Microscale, University of Science and Technology of China, Hefei,
Anhui, 230026, China}
\affiliation{Department of Physics, University of Science and Technology
of China, Hefei, Anhui, 230026, China}
\date{\today}

\begin{abstract}
We investigate the electron spin relaxation of $n$-type InAs quantum
wires by numerically solving the fully microscopic kinetic spin Bloch
equations with the relevant scattering explicitly included.
We find that the quantum-wire
size and the growth direction influence the spin relaxation time by
modulating the spin-orbit coupling. Due to inter-subband scattering
in connection with the spin-orbit interaction, spin-relaxation in
quantum wires can show different characteristics from those in bulk
or quantum wells and can be effectively manipulated by various means.
\end{abstract}
\maketitle

\section{Introduction}
Spintronics continues to attract interest because of potential
applications to information technology, but also because it has
greatly improved our understanding of the role of different
spin-dependent interaction mechanisms, i.e. of
the spin-orbit coupling, for many-electron systems.\cite{spintronics,Fabian}
Several spintronic devices have been proposed that manipulate the
carrier spin via spin-orbit coupling (SOC).\cite{datta,Loss,filter} In
recent years, progress in nanofabrication and growth techniques has
made it possible to produce high-quality quantum wires (QWRs) and
investigate physics in these semiconductor nanostructures.\cite{Quay,
  Ritchie, Hirayama, Wegscheider, Farhangfar, Hardtdegen, Agrawal} 
The energy spectrum of QWR systems with strong SOC has been studied
experimentally\cite{Ulrich2,Fasolino} and theoretically.\cite{Cahay2, Liu,
  Ulrich1,Arakawa, Chang,Bimberg,Vlaev} For $p$-type
QWRs, most of the works concentrate on the energy
spectrum,\cite{Ulrich1, Ritchie} but for $n$-type quantum structures,
many investigations have been performed with the aim of understanding
the electron spin relaxation.\cite{Cahay1, Ridley, Raimondi, Ando,
  Awschalom2, Cheng, VanDerWal, Nitta, Kim, Yoh}

For QWRs, the spin relaxation time (SRT) was measured,\cite{Ando,
  Awschalom2} and calculated in the framework of a single-particle
model\cite{Cahay1,Ridley,Raimondi,Ando} and Monte-Carlo
simulations.\cite{Yoh,Cahay1,VanDerWal,Kim} Recently, Liu \emph{et
  al.}\cite{VanDerWal} investigated the SRT for QWRs in the (110)
crystal direction, but systematic studies of spin relaxation in
$n$-type QWRs are scarce. A general way to obtain the nonequilibrium
spin dynamics and the spin relaxation time in semiconductors
heterostructures has been developed by Wu {\em et
  al.}\cite{Wu1,Weng,Wu5} In this approach, the momentum and spin
dependent distribution functions are calculated using microscopic
kinetic spin Bloch equations (KSBEs), which include the
momentum-dependent Dresselhaus and/or Rashba SOC together with the
effect of the relevant scattering mechanisms.\cite{Wu1,Weng,Wu5} Cheng
\textit{et al.}  applied this approach to study electron spin
relaxation in QWR systems.~\cite{Cheng} However, the influence of
higher subbands was not included in this work. Subsequently, the
effects of higher subbands and their coupling via Coulomb scattering
were shown to important for the spin-relaxation of \emph{holes} in
QWRs\cite{Clv2} and for the case of electrons in quantum
wells.\cite{Oestreich}

In this paper, we study the influence of higher subbands on the spin
relaxation in $n$-type InAs QWRs. This allows us to investigate QWRs
with a wide range of sizes. Especially for larger diameters of the
wires, the spin precession and the spin relaxation are expected to
show different characteristics from narrow ones, as electrons populate
more than one subband and therefore experience different SOCs and
undergo intra-subbband and inter-subband Coulomb scattering. We find
that the inter-subband Coulomb scattering can make an important
contribution to the spin relaxation. We also study the influence of
the growth direction of the QWRs on the spin relaxation. In
particular, we consider QWRs with (001), (110), and (111) growth
directions, and show that the SRT  depends sensitively on the growth
direction, quantum-wire size and the direction of the initial spin
polariazation.

This paper is organized as follows: In Sec.~II we describe our model and the
KSBEs.  Our numerical results are presented in Sec.~III. We conclude
in  Sec.~IV.

\section{Model and  Dynamical Equations}
We model the InAs QWR by a rectangular confinement potential, i.e., we
assume infinitely high barriers at $x = \pm a_x$ and $y = \pm a_y$,
and no  confinement on the $z$ direction. The
Hamiltonian, which describes the electronic
single-particle states in the QWR is then taken to include the
 confinement potential $V_C$, the Rashba term $H_{R}$ and the
Dresselhaus term $H_{D}$  
\begin{equation}
  \label{Luttinger-Hamiltonian}
  H_{e} = \frac{{\mathbf P}^2}{2 m^*} + H_{R} + H_{D} + V_c(\mathbf{ r})~.
\end{equation}
The Rashba and Dresselhaus terms are the two contributions to the
internal $k$-dependent effective field, which leads to the
Dyakonov-Perel' spin dephasing mechanism. The Rashba term
\begin{eqnarray}
  \label{Rashba_100}
    H_{R}(\mathbf{k}) &=& {\gamma^{6c6c}_{41}}
    \mbox{\boldmath$\sigma$\unboldmath} \cdot \mathbf{k}
  \times \mathbf{\mathcal{E}} \nonumber \\ &=& {\gamma^{6c6c}_{41}}[\sigma_x
  (  k_y  {\cal E}_z - 
  k_z {\cal E}_y) + \sigma_y (k_z {\cal E}_x -  k_x 
  \mathcal{E}_z)\nonumber \\ 
  &&\mbox{} + \sigma_z ( k_x  {\cal E}_y -   k_y  {\cal E}_x)]
\end{eqnarray}
 is due to the inversion asymmetry of the crystal structure. The
 Dresselhaus term is different for different growth directions. For a
 (100) InAs QWR, the $x$, $y$ and $z$ axes correspond to the
 [100], [010] and [001] crystallographic directions, respectively, and
 the Dresselhaus term can be written as:~\cite{Winkler_book}
\begin{eqnarray}
  \label{Dresselhaus_100} 
   H^{100}_{D} &=& {b^{6c6c}_{41}} \{ \sigma_x [ k_x ( k_y^2  -
  k_z^2)] + \sigma_y [ k_y (k_z^2 -
   k_x^2 )] \nonumber \\ && \mbox{} + \sigma_z [k_z( k_x^2  -
   k_y^2 )] \} ~.
\end{eqnarray}
For a (110) QWR, the $x$, $y$ and $z$ directions correspond to the
[$\bar{1}$10], [001] and [110] 
crystallographic directions, and we have
\begin{eqnarray}
  \label{Dresselhaus_110} 
  H^{110}_{D} &=& {b^{6c6c}_{41}} \{ \sigma_x [-\frac{1}{2} k_z  (
  k_x^2   -  k_z^2  + 2 k_y^2 )] 
 + 2 \sigma_y  k_x   k_y   k_z \nonumber \\ && \mbox{} + \sigma_z [\frac{1}{2}  
  k_x   (  k_x^2   -  k_z^2  - 2 k_y^2 )]  \}~.
\end{eqnarray}
For a (111) QWR,  the $x$, $y$ and $z$ directions correspond to the
[11$\bar{2}$], [$\bar{1}$10], and [111] crystallographic
directions, and we have
\begin{eqnarray}
  \label{Dresselhaus_111} 
   H^{111}_{D} &=& {b^{6c6c}_{41}} \{ \sigma_x [- \frac{\sqrt{2}}{\sqrt{3}} 
  k_x  k_y  k_z - \frac{1}{2\sqrt{3}} 
  k_y^3   - \frac{1}{2\sqrt{3}}  k_y 
  k_x^2  \nonumber \\ && \mbox{} + \frac{2}{\sqrt{3}}  k_y  k_z^2 -
  \frac{\sqrt{2}}{3}  k_y^2  k_z]  +
  \sigma_y[ \frac{1}{2\sqrt{3}}  k_x^3 \nonumber \\ &&
  \mbox{} +
  \frac{1}{2\sqrt{3}}   k_x   k_y^2 
  -\frac{1}{\sqrt{6}}   k_x^2  k_z  - 
  \frac{1}{\sqrt{6}} k_z ( k_x^2   +  k_y^2
  )] \nonumber \\ &&  \mbox{} + \sigma_z [\frac{\sqrt{3}}{\sqrt{2}} k_x^2
    k_y   - \frac{1}{\sqrt{6}}  k_y^3
   - \frac{2}{3} k_z  k_y^2 ]  \}~.
\end{eqnarray}

As input for the KSBEs we use the basis $\{ \phi_{n_x
  n_yk\sigma} \}$ of single-particle states, which are obtained from
the eigenfunctions of $ \frac{{\mathbf P}^2}{2 m^*} + V_c(\mathbf{ r})$. For a
hard-wall confinement potential which constricts the electrons in the $x$
and $y$ directions on mesoscopic length scales, we can employ the
envelope function approximation.\cite{Haug_Koch}  Thus we
write the single-particle states in the form  
\begin{equation}
  \label{wave_function}
  \phi_{n_x n_yk\sigma}(\mathbf{ r}) = \psi_{n_x,n_y}(x,y)e^{i
    {k}z}\chi_{\sigma} \ ,
\end{equation}
with 
\begin{equation}
  \label{x_y_part}
   \psi_{n_x,n_y}(x,y) = \frac{2}{\sqrt{a_x a_y}} \sin(\frac{n_x\pi
  x}{a_x}) \sin(\frac{n_y\pi y}{a_y})\ ,
\end{equation}
where $\chi_{\sigma} $ denotes the basis vectors in spin space, i.e.,
eigenstates of $\sigma_z$.  In the envelope function approximation,
the effective Hamiltonian acting on the single-particle states is
obtained from Eqs.~(\ref{Rashba_100}-\ref{Dresselhaus_111})  by the
replacements $k_x \to \langle 
\psi_{n_x}|\hat{k}_x |\psi_{n'_x}\rangle \equiv \langle
\hat{k}_x\rangle_{n_x,n_x'}$ and $k_x^2 \to \langle \hat{k}_x^2
\rangle_{n_x,n_x'}$ where $\hat{k}_x = -i\partial/ \partial x$. A
similar replacement is done for $k_y$. For our choice of the
confinement potential, we have in particular
\begin{eqnarray}
  \label{kxkx2}
  && \langle k_x \rangle =  \frac{4i\hbar n_x^{\prime}n_x}{a_x
    [(n_x^{\prime})^2-(n_x)^2]}(1-\delta_{n_x,n_x^{\prime}})\ ,
  \nonumber \\ &&  \langle k_x^2  \rangle = \frac{\hbar^2 \pi^2 n_x^2}{a_x^2}
  \delta_{n_x,n_x^{\prime}} \ .
\end{eqnarray}
with corresponding results for $\langle k_y \rangle$ and $\langle
k_y^2 \rangle$. From Eqs.~(\ref{Dresselhaus_100}) and (\ref{kxkx2})
one can see that when $a_x$ and $a_y$ are 
sufficiently small and only the lowest subband in QWR is important,
only the third term of Eq.~(\ref{Dresselhaus_100}) is not zero, which
means the effective magnetic field contributed by the Dresselhaus term
is along the $z$-direction. On the other hand, the third term in
Eq.~(\ref{Rashba_100}) is zero but the terms which proportional to
$\sigma_x$ and $\sigma_y$ are not zero. This means that the effective
magnetic field contributed by the Rashba term is in the $x$-$y$ plane
when the wire width is sufficiently
small. Nevertheless for larger width of the wire, when higher subbands are
needed, all the terms in Eqs.~(\ref{Rashba_100}) and
(\ref{Dresselhaus_100}) contribute. Moreover, one can
see that when the confinement in the $x$ and $y$ directions is
symmetrical, the third term of Eq.~(\ref{Dresselhaus_100}) is zero,
so that the effective magnetic field contributed by the
Dresselhaus term does not contain any component along the
$z$-direction. As we show in detail below, these differences in the
spin-dependent single-particle states lead to significantly different
behavior of the SRT.

The complete dynamical information about spin-dependent
single-particle properties is contained in the spin-density matrix
$\rho$. Its matrix elements are, in general, defined with
respect to the complete set of quantum numbers $n_x, n_y,k,\sigma$,
but because of the translation symmetry in $z$ direction, $\rho$ is
diagonal in $k$, i.e., $\rho = \rho_{k,ss'}$ with $s = (n_x,
n_y,\sigma)$.

We construct the KSBEs by the non-equilibrium Green function method as
follows:\cite{Wu1,Weng,Haug} 
\begin{equation}
  \label{Bloch_eq}
  \frac{\partial\rho}{\partial t} =  \frac{\partial\rho}{\partial
    t}\Big|_{\rm {coh}} +
  \frac{\partial\rho}{\partial t}\Big|_{\rm {scat}}. 
\end{equation}
The coherent terms can be written as
\begin{eqnarray}
  \label{coh}
  \frac{\partial\rho_{k}}{\partial t}\Big|_{\rm coh} = - i \Big[
  \sum_{\mathbf{Q}} V_{\mathbf{Q}}  
  I_{\mathbf{Q}} {\rho}_{k-q} I_{-\mathbf{Q}} ,  \rho_k \Big]
  -i \Big[ H_{e}(k),  \rho_k   \Big], 
\end{eqnarray}
where $[A,B] = AB - BA$ denotes the commutator, and $\mathbf{Q} \equiv
(q_x, q_y, q)$.  $I_{\mathbf{Q}} $ is a matrix in $(s,s')$ space and
can be considered as a form factor. Its definition reads 
\begin{eqnarray}
  I_{\mathbf{Q},s_1,s_2} &=& \langle s_1 | e^{i\mathbf{Q} \cdot
   \mathbf{r}} | s_2 \rangle  
 \nonumber \\
  &=& \delta_{\sigma_1,\sigma_2}F(m_1,m_2,q_y,a_y)F(n_1,n_2,q_x,a_x)
\end{eqnarray}
where
\begin{widetext}
  \begin{equation}
    F(m_1 , m_2,q,a)= 2 i a q [e^{i a q} \cos{\pi (m_1 -
      m_2)} -1 ] 
    \left[\frac{1}{\pi^2(m_1 - m_2)^2 -
        a^2 q^2}
      - \frac{1}{\pi^2(m_1 + m_2)^2 - a^2 q^2}\right].
  \end{equation}
\end{widetext}
The first term in Eq.~(\ref{coh}) is the Coulomb Hartree-Fock term,
and the second term is the contribution from the single-particle
Hamiltonian, i.e., Eq.~\eqref{Luttinger-Hamiltonian} in $(s,s')$
space, which includes the spin-orbit coupling terms. 
For small spin polarization, the contribution from the Hartree-Fock
term in the coherent term is negligible\cite{Weng,schu} and the
coherent spin dynamics 
is essentially due to the spin precession around the effective
internal fields described by Eqs.~(\ref{Rashba_100}-\ref{Dresselhaus_111})

The scattering contributions to the dynamic equation of the
spin-density matrix include scatterings between electrons and
nonmagnetic impurities, electrons and phonon, and electron and electron
scatterings:
\begin{widetext}
 \begin{eqnarray}
   \frac{\partial \rho_{{k}}}{\partial t}\Big|_{\rm scat} &=&
   \frac{\partial \rho_{{k}}}{\partial t}\Big|_{\rm im} + 
   \frac{\partial \rho_{{k}}}{\partial t}\Big|_{\rm ph} +
   \frac{\partial \rho_{{k}}}{\partial t}\Big|_{\rm ee} \ , \nonumber \\ 
   \frac{\partial \rho_{{k}}}{\partial t}\Big|_{\rm im} &=& \pi N_i \sum_{\mathbf{
       Q},s_1,s_2} |U^i_{\mathbf{Q}}|^2
   \delta(E_{s_1,k-q} - E_{s_2,k})  I_{\mathbf{Q}} 
   [(1-{ \rho}_{k-q})T_{s_1}
   I_{-{\mathbf{Q}}} T_{s_2} {
     \rho}_k -
   \rho_{k-q} T_{s_1} I_{-{\mathbf{Q}}} T_{s_2} (1-\rho_k)]  + h.c.\ , \nonumber  \\
   \frac{\partial \rho_{{k}}}{\partial t}\Big|_{\rm ph} &=&  \pi \sum_{{\mathbf{Q}},s_1,s_2,\lambda}
   |M_{{\mathbf{Q}},\lambda}|^2 
   I_{\mathbf{Q}} \{
   \delta(E_{s_1,k-q} - E_{s_2,k} +
   \omega_{{\mathbf{Q}},\lambda} ) [(N_{{\mathbf{Q}},\lambda} +1)(1-{
     \rho}_{k-q})  T_{s_1}I_{-{\mathbf{Q}}} T_{s_2} { \rho}_k \nonumber \\
   &&\mbox{} - N_{\mathbf{
       Q},\lambda} \rho_{k-q}  T_{s_1} I_{-{\mathbf{Q}}} T_{s_2}(1-\rho_k)]
   +\delta(E_{s_1,k-q} - E_{s_2,k} -
   \omega_{{\mathbf{Q}},\lambda} ) [N_{{\mathbf{Q}},\lambda}(1-{
     \rho}_{k-q})  T_{s_1}I_{-{\mathbf{Q}}}
   T_{s_2}{ \rho}_k   \nonumber \\ 
 &&\mbox{}- (N_{\mathbf{
       Q},\lambda}+1) \rho_{k-q}  T_{s_1}I_{-{\mathbf{Q}}} T_{s_2}(1-\rho_k)]
   \}+ h.c. \ ,  \nonumber \\
   \frac{\partial \rho_{{k}}}{\partial t}\Big|_{\rm ee} &=&  \pi \sum_{{\mathbf{Q}},k^{\prime}}
   \sum_{s_1,s_2,s_3,s_4} V_{\mathbf{Q}}^2   \delta(E_{s_1,{
       k}-q} - E_{s_2,k} + E_{s_3,k^{\prime}} -
   E_{s_4, k^{\prime} -q})I_{\mathbf{Q}}
  \nonumber \\
 &&\mbox{}\times \{
   (1-{\rho}_{k-q})T_{s_1}
   I_{-{\mathbf{Q}}} T_{s_2}{   \rho}_k
   \mbox{Tr}[(1-\rho_{k^{\prime}})T_{s_3}
   I_{\mathbf{Q}} T_{s_4}\rho_{k^{\prime} -q}
   I_{-\mathbf{ Q}}]  \nonumber \\
 &&\mbox{}- \rho_{\mathbf{
       k}-q} T_{s_1} I_{-\mathbf{ Q}} T_{s_2}(1-\rho_k)
   \mbox{Tr}[\rho_{k^{\prime}}T_{s_3}
   I_{\mathbf{Q}}  T_{s_4} (1-\rho_{
       k^{\prime}-q})
   I_{-\mathbf{ Q}} ]\}  + h.c. \ \ ,
 \label{scat}
 \end{eqnarray}
 \end{widetext}
 in which $T_{s_1,s,s'} =  \delta_{s_1, s} \delta_{s_1,s'}$. 
 The statically screened Coulomb potential in the random-phase
 approximation (RPA) reads~\cite{Haug}
 \begin{equation}
   \label{V_q}
   V_{q} = {\sum_{q_x,q_y}v_{Q}|I_{\mathbf{ Q}}|^2}/{\kappa(q)},
 \end{equation}
 with the bare Coulomb potential $v_{Q} = 4 \pi e^2 / Q^2 $ and 
 \begin{equation}
   \label{RPA}
   \kappa(q) = 1- {\sum_{q_x,q_y}v_{Q}|I_{\mathbf{ Q}}|^2} \sum_k
   \frac {f_{k+q} - f_k}{\epsilon_{k+q} - \epsilon_k}.
 \end{equation}
 In Eq.~(\ref{scat}), $N_i$ is the density of impurities, and $
 |U^{i}_{\mathbf{Q}}|^2 $ is the impurity potential. Further, $|M_{\mathbf{
     Q},\lambda}|^2$ and $N_{\mathbf{ Q},\lambda} =
 [\mbox{exp}(\omega_{\mathbf{ Q},\lambda} /k_B T) - 1]^{-1}$ are the
 matrix element of the electron-phonon interaction and the Bose
 distribution function, respectively. The phonon energy spectrum for
 phonon mode $\lambda$ and wavevector $\mathbf{Q}$ is denoted by
 $\omega_{\mathbf{ Q},\lambda}$. For the electron-phonon scattering, we
 include the electron LO-phonon and electron AC-phonon scattering, for
 which the explicit expressions can be found in
 Refs.~\onlinecite{Weng} and \onlinecite{Zhou}. Note that, due to the
 weakness of the 
 SOC, the energy dispersions $E_{s,k}$ in the scattering terms,
 Eq.~(\ref{scat}), are taken from the diagonalization of $
 \frac{{\mathbf P}^2}{2 m^*} + V_c(\mathbf{ r})$ and do not include
 the SOC, as in 
 Ref.~\onlinecite{Cheng2}. This is different from our previous work in
 $p$-type QWR systems.~\cite{Clv2} For holes, the SOC is stronger
 because holes are derived from $p$-orbitals, which experience the
 spin-orbit interaction directly, and it is therefore necessary to
 include the SOC in the single-particle energy
 dispersions.\cite{Cheng2,Clv2,hans} 

\section{Numerical Results}

We numerically solve the KSBEs for the spin density
matrix $\rho$, from which we obtain the dynamics of the average spin
for electrons with momentum $k$ via
\begin{equation}
\langle\mathbf{S}\rangle _k(t) = \sum_{n_x, n_y,\sigma} \rho_{k,n_x, n_y,\sigma,n_x,
  n_y,\sigma}(t) \langle \sigma| \mathbf{s}|\sigma\rangle \ ,
\end{equation} 
where $\mathbf{s}$ is the single-particle spin operator.  The SRT $\tau$
can then be defined by an exponential fit to the envelope of the $z$
component of the average spin of the ensemble of
electrons:~\cite{Weng}
\begin{eqnarray}
  \label{sum_S}
\langle S\rangle_z = \sum_k (\langle\mathbf{S}\rangle_{k})_z(t).
\end{eqnarray}

In all the numerical results, we include the electron-phonon and
electron-electron scattering. As initial condition we assume
a spin polarization along the $z$-direction with a small initial spin
polarization $ P= (2/\hbar)\langle S\rangle_z/N_e $ where $N_e$ is the
total electron density.

\begin{table}[ht]
\caption{Material parameters used in the calculation (from
   Ref.~\onlinecite{parameter} unless otherwise specified).}
\begin{center}
\begin{tabular}{lllllllllll}
  \hline \hline
  $\kappa_{\infty}$ & $12.25$ &\mbox{}&\mbox{}&\mbox{}&\mbox{}&\mbox{}&\mbox{}&\mbox{}& $\kappa_0$ & $15.15$ \\
  $m_e/m_0$ & $0.023$ &\mbox{}&\mbox{}&\mbox{}&\mbox{}&\mbox{}&\mbox{}&\mbox{}& $\Omega_{\mbox{\tiny LO}}$~(meV)
   & 27.0 \\
  $v_{sl}$~(km/s) & 4.28 &\mbox{}&\mbox{}&\mbox{}&\mbox{}&\mbox{}&\mbox{}&\mbox{}& 
  $v_{st}$~(km/s) & 1.83 \\
  $b^{6c6c}_{41}$~(e\AA$^3$) & 27.18$^a$ &\mbox{}&\mbox{}&\mbox{}&\mbox{}&\mbox{}&\mbox{}&\mbox{}&
  $\gamma^{6c6c}_{41}$~(e\AA)   & 117.1$^a$ \\ 
  $\symbol{'4}$~(eV) & $5.8$ &\mbox{}&\mbox{}&\mbox{}&\mbox{}&\mbox{}&\mbox{}&\mbox{}& $e_{14}$~(V/m) &
  $0.35\times 10^9$\\ 
  $\Delta_{0}$~(eV) & 0.38 &\mbox{}&\mbox{}&\mbox{}&\mbox{}&\mbox{}&\mbox{}&\mbox{}& $E_g$~(eV) &  0.414 \\
  \hline \hline
\end{tabular}
\end{center}
\label{parameter}
\hspace {-13pc} $^a$ Ref.~[\onlinecite{Winkler_book}].
\end{table}

\begin{figure}[thb]
  \begin{center}
    \includegraphics[width=8.5cm]{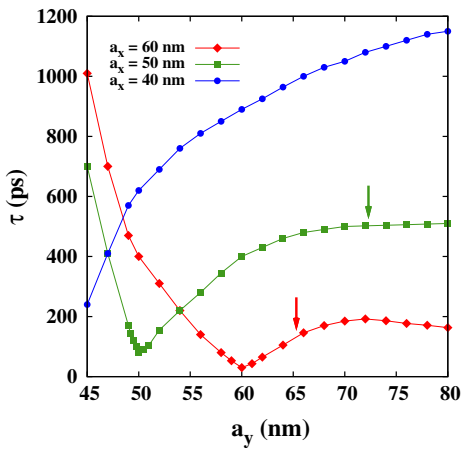}
  \end{center}
  \caption{SRT $\tau$  {\em vs.} the QWR width in $y$ direction, $a_y$,
    for (100) QWRs at different $a_x$. The electron density is $N = 4 \times 10^5$~cm$^{-1}$
    and $T = 100$~K. The arrows mark the densities at which the electron
    populations in the second and higher subbands are
    approximately 30~\%.}
\end{figure}

\subsection{Influence of the wire size}

In Fig.~1 we plot the SRT as a function of the QWR width in $y$
direction, $a_y$, for various $a_x$. We choose (100) QWRs at a lattice temperature
of $T = 100$~K and a total electron density of $N_e = 4 \times
10^5$~cm$^{-1}$. For a small QWR of width $a_x = 40$~nm, the SRT
increases monotonously with $a_y$ because for a small QWR the spacing
of the subbands is large, so that for the present conditions only the
lowest subband in QWR is appreciably populated. Therefore, as
discussed in Sec.~II, the effective magnetic field due to the
Dresselhaus term contains a longitudinal component, $B^D_z(k)$, that
keeps the electronic spins aligned and thus inhibits spin precession,
which is mainly due to the Rashba term, thereby effectively reducing
the spin relaxation.\cite{Fabian}
An interesting effect arises because $B^D_z(k)$ is proportional to
$(\langle k_x^2 \rangle - \langle k_y^2 \rangle)$, which disappears
when the wire widths, and therefore in the present model also the
confinement wave functions, in $x$ and $y$ directions are identical.
For vanishing $B^D_z(k)$, i.e., for $a_x= a_y$, the SRT reaches a
minimum of several 10~ps for the QWRs considered here. Changing the
wire size, leads to increasing $(\langle k_x^2 \rangle - \langle k_y^2
\rangle)$ and therefore to increasing SRT. This is clearly illustrated
for the $a_x = 50$~nm case. When $a_x = 60$~nm, one can also find a
minimum of SRT at $a_x = a_y = 60$~nm for the reason described
above. However, one finds that the SRT decreases with $a_y$ again when
$a_y > 70$~nm. This is because the electrons start to populate higher
subbands when the wire width increases. Then the effective magnetic
field formed by the Dresselhaus term contains not only a longitudinal
component but also a transverse component, $B^D_{\|}(k)$, which
couples different subbands. This coupling has the same effect as a
spin precession, and therefore tends to make the SRT
shorter. Moreover, the contribution of the Dresseslhaus term becomes
as important as the contribution of the Rashba term. When $a_y >
70$~nm, $B^D_{\|}(k)$ dominates over $B^D_{z}(k)$ due to the
increasing $a_y$, and a faster spin relaxation results.  The
different minima of the SRT at $a_x = a_y = 60$~nm and $a_x = a_y =
50$~nm are likely due to the different energy gaps between different
subbands in the two cases. 

\subsection{(110) and (111) QWRs}

\begin{figure}[thb]
  \begin{center}
    \includegraphics[width=8.5cm]{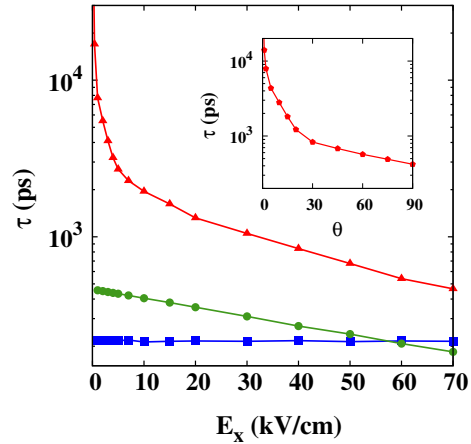}
  \end{center}
  \caption{SRT $\tau$  {\em vs.} $E_x$ for (110) QWRs 
at $T = 50$~K and $N_e = 4 \times
    10^5$~cm$^{-1}$. {$\blacktriangle$}:  $a_x = a_y = 30$~nm with an
    initial spin polarization along the $x$-direction; $\blacksquare$:
    $a_x = a_y = 30$~nm with an initial spin polarization along
    the $y$-direction; {$\bullet$}: $a_x = a_y = 50$~nm with an
    initial spin polarization along the $x$-direction.} 
\end{figure}

The results of the previous subsection showed that the geometry of the
QWR has a pronounced effect on the SRT because it influences the
different contributions to the SOC. We now investigate QWRs with
different growth directions, and start with the case of (110)
QWRs. The SOC for (110) QWR is quite different from the SOC for (100)
QWR as shown in Eq.~(\ref{Dresselhaus_100}) and
Eq.~(\ref{Dresselhaus_110}). First we only consider the case of a
narrow wire with $a_x = a_y = 30$~nm, for which the electronic
population is mainly in the lowest subband. In the presence of an
electric field of the form $(E_x,E_y,0)$, the relevant contributions are
\begin{eqnarray}
  \label{Rashba_110_1band}
  &&  H^{110}_{R} = {\gamma^{6c6c}_{41}}[-\sigma_x {
    E}_y k_z + \sigma_y E_x k_z]~, \\
  \label{Dresselhaus_110_1band} 
  &&  H^{110}_{D} =  -\frac{1}{2} {b^{6c6c}_{41}} \sigma_x k_z [ \langle
  k_x^2 \rangle  -  k_z^2  + 2\langle k_y^2 \rangle ] ~.
\end{eqnarray}
The effective magnetic field formed by the Dresselhaus term is along
the $x$-direction, which corresponds to the [$\bar{1}$10]
crystallographic direction, and the effective magnetic field formed by
the Rashba term is in the $x$-$y$ plane. If the direction of the total
effective magnetic field formed by the SOC is tuned to be exactly the
direction of the initial spin polarization, then one can expect an
extremely long SRT as pointed out in
Refs.~\onlinecite{VanDerWal} and \onlinecite{Cheng2}. For the QWRs
considered in this 
paper, we can study the physics that gives rise to this effect in the
following way: We take the initial spin polarization to be along the
$x$-direction. In Fig.~2 we plot the SRT as a function of $E_x = E
\cos{\theta}$ for temperature $T = 50$~K, electron density $N_e = 4
\times 10^5$~cm$^{-1}$, and wire geometry $a_x = a_y = 30$~nm. The SRT
decreases with $E_x$, which is a measure of the effective magnetic
field along $y$-direction. As expected, a very long SRT results when
$E_x$ is very small, even though the effective magnetic field along
the $x$-direction is not zero.  In comparison, for an initial spin
polarization along the $y$-direction a much shorter SRT is obtained,
which hardly changes with $E_x$. This is because the polarized
electronic spins precess around the $x$-direction and this precession
is not influenced strongly by the $y$-component of the effective
magnetic field. In the inset of Fig.~2, we also show the dependence of
the SRT on the angle $\theta$, which the electric field in the $x$-$y$
plane forms with the $x$ axis. Assuming $E= 10$~kV/cm and an initial
spin polarization along the $x$-direction, it is found that the SRT
decreases with $\theta$. For small $\theta$, i.e., for effective
magnetic fields close to the $x$-direction, the SRT goes up, in
agreement with the previous discussion. The result for a larger wire
size, $a_x = a_y = 50$~nm are also plotted in Fig.~2. In this case
the SRT is never longer than 1~ns, even when $E_x$ is very small
because now electrons populate higher subbands, and the effective
magnetic field formed by the Dresselhaus term contains not only the
longitudinal component but also a transverse component. 

\begin{figure}[thb]
  \begin{center}
    \includegraphics[width=8.5cm]{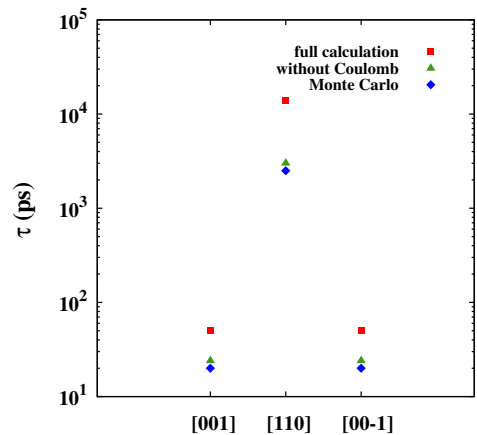}
  \end{center}
  \caption{Our results compare with the results obtained by
    Monte-Carlo simulations with different wire directions at $T =
    5$~K, $N_e = 4 \times 
    10^8$~cm$^{-1}$, $N_i = 3.17 \times 10^6$~cm$^{-1}$, and $a_x =
    a_y = 1~\mu$m.} 
\end{figure}

We compare the SRT obtained by our calculation with Monte-Carlo
results from Ref.~\onlinecite{VanDerWal} in Fig.~3. for a
temperature of $T = 5$~K. The total electron density and the impurity
density are taken to be $N_e = 4 \times 10^8$~cm$^{-1}$ and $N_i =
3.17 \times 10^6$~cm$^{-1}$, and the wire geometry is $a_x = a_y =
1~\mu$m. The results without the Coulomb scattering are actually very
close to those obtained by Monte-Carlo simulations. However, when the
Coulomb scattering is included, much longer SRTs result. This result
underscores the importance of Coulomb scattering for spin relaxation
in QWRs.

\begin{figure}[thb]
  \begin{center}
    \includegraphics[width=8.5cm]{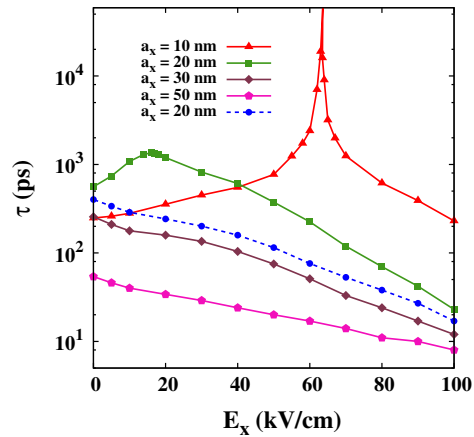}
  \end{center}
  \caption{SRT $\tau$  {\em vs.} $E_x$ for different wire
    sizes $a_x = a_y$ at $T = 50$~K and $N_e = 4 \times
    10^5$~cm$^{-1}$. The growth direction of the QWR is along the
      [111] crystallographic direction. The solid curves are the
    results with the initial 
    spin polarization along $z$-direction and the dashed curve is the
   result with the initial spin polarization along $x$-direction.} 
\end{figure}

For the quantitative analysis of the SRT for another growth direction
we choose (111) QWRs. Here, we again consider the case of small wire
width and low 
temperature first. The electric field is again taken to be
$(E_x,E_y,0)$, so that the SOC can be written as
\begin{eqnarray}
  \label{Rashba_111_1band}
   H^{111}_{R} &=& {\gamma^{6c6c}_{41}}[-\sigma_x {
    E}_y k_z + \sigma_y E_x k_z]~, \\
  \label{Dresselhaus_111_1band} 
   H^{111}_{D} &=& {b^{6c6c}_{41}} \{ -  \frac{\sqrt{2}}{3} \sigma_x k_z  \langle
  k_y^2 \rangle  -
  \frac{1}{\sqrt{6}}\sigma_y k_z (\langle k_x^2 \rangle  + \langle k_y^2
  \rangle) \nonumber \\ && \mbox{} -\frac{2}{3} \sigma_z 
    k_z \langle k_y^2 \rangle  \}~.
\end{eqnarray}
Similar to the case of (110) QWRs, we expect very long SRTs if the
total effective magnetic field points into the direction of the
initial spin polarization, at least for narrow wires. For a numerical
example of this effect, we choose $E_y$ such that $\gamma^{6c6c}_{41}
E_y + (\sqrt{2}/3)b^{6c6c}_{41} \langle k_y^2 \rangle = 0 $ for a
small wire with $a_x = a_y = 10$~nm, so that the $x$ component of the
total effective magnetic field is zero. For an initial spin
polarization along the $z$-direction, which corresponds to the [111]
crystallographic direction, we plot the SRT as a function of $E_x$ in
Fig.~4 at $T = 50$~K and $N_e = 4 \times 10^5$~cm$^{-1}$. Fig.~4 shows
that when $a_x = a_y = 10$~nm, there is a pronounced maximum of the
SRT at $E_x = 70$~kV/cm, which fulfils the relation
${\gamma^{6c6c}_{41}} E_x + \frac{1}{\sqrt{6}}{b^{6c6c}_{41}}(\langle
k_x^2 \rangle + \langle k_y^2 \rangle ) \approx 0$.  Consequently, for
this field strength, the direction of the total effective magnetic
field is exactly along the direction of the initial spin polarization
and this leads to a very long SRT. However, this effect much reduced
for larger wire cross sections. When $a_x = a_y = 20$~nm, there is
still a maximum of the SRT, but the maximum is much less pronounced
than for the smaller wire, because now ${\gamma^{6c6c}_{41}} E_y +
\frac{\sqrt{2}}{3}{b^{6c6c}_{41}} \langle k_y^2 \rangle $ remains
finite for all field strengths as $\langle k_y^2 \rangle $ is changed.
This trend continues for wire sizes of $a_x = a_y = 30$~nm and $a_x =
a_y = 50$~nm. Finally, we analyze the the case of $a_x = a_y =
20$~nm with the initial spin polarization along $x$-direction: Here
one also does not obtain a maximum of the SRT because the direction of
the effective magnetic field is no longer identical to the direction
of the initial spin polarization.

\subsection{Doping and temperature dependence}

\begin{figure}[thb]
  \begin{center}
    \includegraphics[width=8.5cm]{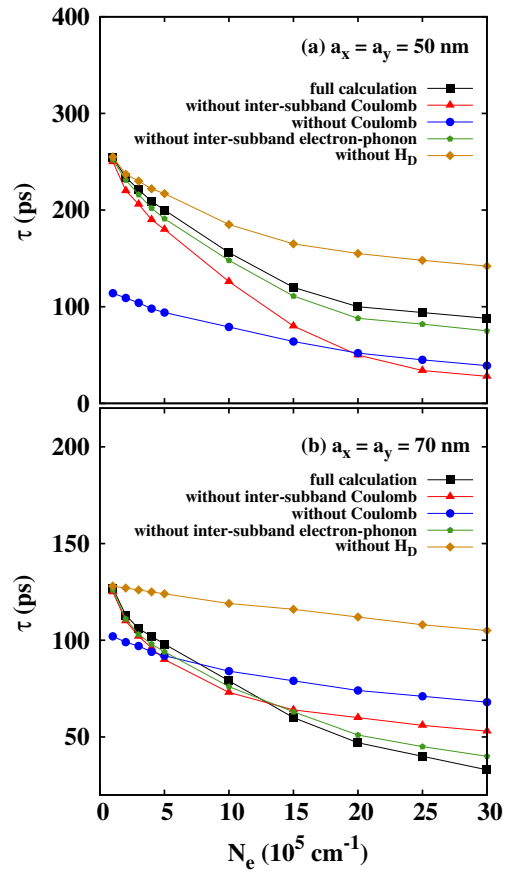}
  \end{center}
  \caption{SRT $\tau$ {\em vs.} the electron density for (100)
      QWRs at different wire sizes. (a) $a_x = a_y =
      50$~nm; (b) $a_x = a_y =
    70$~nm (b). $T = 100$~K.}
\end{figure}

Since the population of higher subbands has been shown to play an
extremely important role for the SRT, we next analyze how the
electronic population can be changed without using a different wire
geometry or growth direction, namely by varying the doping density
and/or temperature.  In Fig.~5(a) we plot the SRT as a function of
$N_e$ for a (100) QWR of size $a_x = a_y = 50$~nm, and $T= 100$~K. For
the full calculation, the SRT decreases with $N_e$ because more
electrons are present at higher momenta and in higher subbands, so
that the effective magnetic fields experienced by these electrons are
larger. By the DP mechanism, this results in a faster spin
relaxation.\cite{Fabian} To investigate how scattering affects the
spin relaxation, we first switch off inter-subband
\emph{electron-phonon} scattering, 
with the result that the SRT
becomes shorter. When we switch off the inter-subband
\emph{electron-electron} scattering, the SRT also becomes shorter.
This somewhat counterintuitive effect of scattering on the spin
dephasing results because the influence of scattering is different for
different regimes, i.e., for strong and weak scattering: If we define
$\Omega= H_R + H_D$ and $\tau^*$ to be the effective momentum
relaxation time, then $|{\Omega}| \tau^*$ is typically much smaller
than 1 for the densities considered here ($|\Omega| \tau^* = 0.03$ at
$N_{e} = 10 \times 10^{5}$\ cm$^{-2}$ if $\tau^* $ contains only the
contribution from the inter-subband electron-electron
scattering). This means we are in the strong scattering regime where
the motional narrowing picture \emph{qualitatively} describes the
dependence of the SRT on the momentum scattering time as SRT $\approx
1/\tau^*$. (Microscopically this means that the scattering is strong
enough to prevent significant deviations from isotropic electron
distributions, which are caused by the anisotropic SOC contributions.)
Thus switching off scattering contributions effectively leads to a
longer momentum relaxation time, and therefore to shorter SRTs, while
still remaining in the strong scattering regime. 
Further, our
calculation shows that the inter-subband electron-electron scattering
is more important than the inter-subband electron-phonon
scattering. When we further exclude the intra-subband
electron-electron scattering, it is found that the SRT becomes shorter
when $N_e < 20 \times 10^5$~cm$^{-1}$, but becomes longer when $N_e >
20 \times 10^5$~cm$^{-1}$. Here the most dominant scattering process
is suppressed, so that with increasing $N_e$ the motional narrowing
regime is left, because electrons occupy states at higher momenta and
higher subbands where the SOC contributions rapidly increase. For
higher densities, the weak scattering regime is reached where the
anisotropy of the SOC contributions becomes dominant. Additional
scattering leads to a more efficient dephasing, so that the SRT
increases when we switch off scattering contributions.

We also plot the case of larger wire width with $a_x = a_y = 70$~nm
in Fig.~5(b). Compared to the case of Fig.~5(a), the electrons are
populating higher subbands, so that the Dresselhaus term becomes more
important. Therefore, the band structure anisotropy will become
important at smaller $N_e$. Fig.~5(b) shows that the result of the
full calculation intersects the one without the electron-electron
scattering at $N_e \approx 8 \times 10^5$~cm$^{-1}$. For this wire
geometry, the result without the inter-subband electron-electron
scattering also crosses the full calculation. To see in more detail
how the strength of anisotropic SOC contributions affects the spin
relaxation, we also plot the result without the Dresselhaus term. For
this somewhat artificial case, the SRT becomes much longer. This
indicates that the contribution of the Dresseslhaus term is very
important for this wire width in contrast to the case of smaller
 wire width in Fig.~5(a), because the transverse component of the
  effective magnetic field contributed by the Dresselhaus term is
important only for the large wire width when higher subbands are populated.

\begin{figure}[thb]
  \begin{center}
    \includegraphics[width=8.5cm]{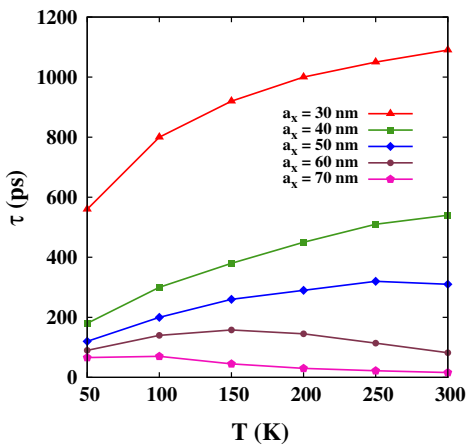}
  \end{center}
  \caption{SRT $\tau$ {\em vs.} the temperature for (100) QWRs at different wire
    sizes $a_x = a_y$ at carrier density $N = 4 \times 10^5$~cm$^{-1}$.}
\end{figure}

Finally we look at the temperature dependence of the SRT at different
wire widths in Fig.~6. We choose (100) QWRs at 
a representative electron density of
$N_e = 4 \times 10^5$~cm$^{-1}$. The temperature affects the SRT in
two ways: For the smaller wires $a_x = a_y \le 50$~nm the carriers
are confined in the lowest subband without populating higher
$k$-states where the anisotropic SOC contributions to the
bandstructure become much stronger. In this strong scattering regime,
the behavior can qualitatively be explained by the motional narrowing:
Increasing temperature leads to enhanced scattering with phonons and
electrons with higher kinetic energies, so that the effective carrier
lifetime becomes shorter, and the SRT increases. Around $a_x = a_y =
60$~nm and $T > 150$~K this behavior crosses over to the weak
scattering limit, in which higher subbands are populated, the
precession frequencies around the effective internal fields become
higher, and additional scattering leads to a more efficient dephasing,
so that the SRT decreases with $T$.

\section{Conclusion}

In conclusion, we have investigated the spin relaxation of electrons in
$n$-type InAs QWRs. The SRT is calculated by numerically solving the
microscopic KSBEs including multiple
subbands. The inclusion of higher subbands allows us to investigate
QWRs larger QWRs than in Ref.~\onlinecite{Cheng}, and we find that the
quantum-wire size influences the spin relaxation time via the
SOC: The Dresselhaus term contains a longitudinal
contribution to the internal effective magnetic field, which can
effectively reduce the spin precession, and thereby the spin
relaxation. It also contains a transverse component, which provides a
contribution to the spin precession involving different subbands and
enhances spin relaxation.  When the wire width is small and only the
lowest subband in the QWR is important, the longitudinal term is
dominant. When the wire width is large and higher subbands are
populated, the transverse contribution dominates over the longitudinal
one, and a faster spin relaxation results. We also studied different
growth directions for QWRs. We show that one can obtain long spin
relaxation time by optimizing the growth direction, quantum-wire width
and the direction of the initial spin polariazation. Further, we
investigated how the details of the microscopic scattering mechanisms
and the spin-orbit effects in the band structure affect the spin
relaxation. The population of higher subbands was found to have
decisive influence on the behavior of the SRT. For
instance, if the geometry and external conditions are such that higher
subbands become populated, the dependence of the SRT
on temperature is reversed because the motional narrowing regime is
left.

\begin{acknowledgments}
This work was supported by the Natural Science Foundation of China
under Grant No.~10725417, the National Basic Research Program of
China under Grant No.~2006CB922005 and the Knowledge Innovation
Project of the Chinese Academy of Sciences. We have also benefitted from a
German-Chinese cooperation grant from the Bosch Foundation. One of the
authors (C.L.)  thanks J. H. Jiang for many discussions.
\end{acknowledgments}

\end{document}